\documentclass[aps,twocolumn,superscriptaddress,showpacs,prl]{revtex4}

\usepackage{graphicx,amsfonts,amsmath,color,amsbsy,amssymb}
\usepackage{epstopdf}

\begin{document}
\title{Optonanofluidics: Modelling fluid flow through surfactant-modified liquid tethers by laser beams}

\author{Joshua A. Bull}
\affiliation{Department of Mathematical Sciences, Durham University,Stockton Road, Durham, DH1 3LE, UK.}

\author{Alex L. Hargreaves}
\affiliation{Department of Chemistry, Durham University, Stockton Road, Durham, DH1 3LE, UK.}

\author{Colin D. Bain}
\affiliation{Department of Chemistry, Durham University, Stockton Road, Durham, DH1 3LE, UK.}

\author{Buddhapriya Chakrabarti}
\email{b.chakrabarti@sheffield.ac.uk}
\affiliation{Department of Physics and Astronomy, University of Sheffield, Hounsfield Road, Sheffield, S3 7RH, UK.}

\date{\today}
\begin{abstract}
When a surfactant-stabilised oil droplet with an ultralow interfacial tension is trapped in the focus of two laser beams and pulled apart (by moving the laser beams) a configuration of two droplets connected by a thin tether of oil results. The tether radius depends on the ratio of the bending modulus to the renormalized interfacial tension, which takes into account the spontaneous curvature of the interface. The force exerted by the tether on the droplets is shown to be asymmetric with respect to the phase inversion temperature of the emulsion, in agreement with experiment. Fluid can be pumped from one droplet to the other via the tether by increasing the optical pressure on one droplet. The flow is a combination of Poiseuille flow within the thread of oil and the external flow around a rigid cylinder, with the surface velocity determined by tangential stress balance. For typical viscosities of oils and the continuous aqueous medium, flow is predominantly in the external medium. The normal stress balance leads to a variation in the radius of the thread with distance. The radius is shown to decrease approximately linearly with a slope proportional to the volumetric flow rate through the tether. For a tether of a given length, there is therefore an upper limit to the flow rate that can be generated by pumping with optical traps.
\end{abstract}
\maketitle

\textit{Introduction:}
Fluid flow through rigid pipes under an imposed pressure gradient is a textbook problem~\cite{b:Landau1959}. With the no-slip boundary condition imposed (\textit{i.e.} fluid velocity being zero at the walls) a Poiseuille flow profile with the flow rate proportional to the fourth power of the pipe radius and inversely proportional to its length is established~\cite{b:Landau1959}. The problem is more complex when fluid flow occurs through flexible pipes where the no-slip boundary condition cannot be enforced. An example of such a situation is flow of blood through veins and arteries~\cite{b:Thirrett2008}. While the problem is resolved by balancing normal and tangential stresses at the boundaries, the coupling between shape and fluid flow leads to novel phenomena including nonlinear pressure-drop/flow-rate relations, self-excited oscillations of single-phase flow at high Reynolds number, capillary-elastic instabilities of two-phase flow at low Reynolds number, \textit{etc.}~\cite{r:Jensen2004,p:Olmsted1997} and it continues to remain an active area of research. 

A similar shape-flow coupling arises in context of giant unilamellar vesicles connected by extruded nanometer sized lipid tubules. In these systems, fluids can be transported through the tether from one vesicle to another via microinjection, by inducing surface tension gradients through squeezing one of the vesicles or by adding lipids~\cite{p:Dommersness2005,p:Joanny2006}. When one of the vesicles is squeezed with a microneedle a surface tension gradient sets up over a few seconds following which the tether relaxes to a stable equilibrium shape~\cite{p:Dommersness2005}. The time scale of relaxation and the steady state shape equations of the tether were calculated using force balance across the tether boundary assuming a flow ansatz~\cite{p:Dommersness2005}. 
  
\begin{figure}
\includegraphics[scale=0.24]{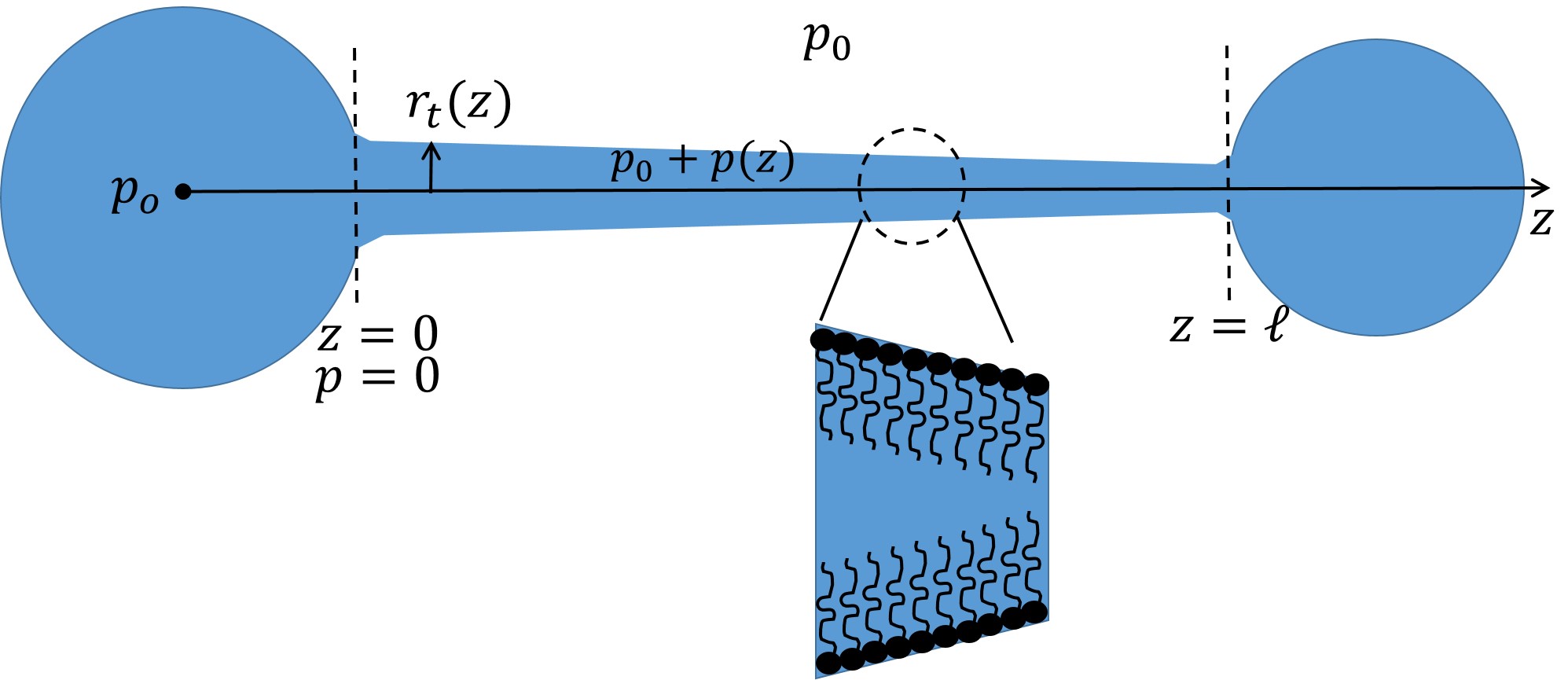}
\caption{Schematic figure showing a tether formed by pulling apart a droplet trapped in the focus of two laser beams with a magnified section of the tether showing a surfactant monolayer.}
\label{fig:Infinite_Cylinder}
\end{figure}

The mechanical stability and flow through threads of oil of nanometric thickness connecting oil droplets in an oil-in-water emulsion is the focus of the present work.  If surfactants are used to reduce the oil-water interfacial tension to ultralow values ($\lesssim 10^{-5} N m^{-1}$), optical traps can be used to control the shape of the oil droplets~\cite{p:Bain2006, p:Chakrabarti2014}. If two optical traps are placed in a single oil droplet (a few microns in diameter) and then pulled apart, a configuration of two droplets connected by a thin tether results (Figure \ref{fig:Infinite_Cylinder}~\cite{p:Bain2011}). This tether is mechanically stable and in the absence of pumping can be made arbitrarily long. The radiation forces from the laser beams of the optical traps results in a negative hydrostatic pressure within the terminal droplets. Liquid can be pumped from one droplet to the other through the tether by variation of the laser intensities in the two traps.  In contrast to the case of the lipid tubules, the surfactant equilibrates rapidly with the interface on the timescale of surface deformations and therefore the interfacial tension can be considered independent of the rate or extent of deformation. Thermal Marangoni effects may arise from the temperature-dependence of the interfacial tension - the optical traps cause a small amount of heating~\cite{p:Hargreaves2015} - but we do not consider this effect here.

In what follows we solve the shape-flow problem for this experimental system. Our main results can be summarised thus: first, we show that the radius of the tether has the expected square root dependence on the bending modulus over the surface tension provided that the surface tension is renormalized to incorporate the energetic cost of flattening the interface. This renormalisation provides a simple explanation for the non-monotonic variation of the interfacial tension in the vicinity of microemulsion phase transitions when temperature or salt concentration is varied [10]. It also explains the asymmetry observed experimentally in the forces exerted by the oil tethers on the terminal droplets~\cite{p:Bain2011}. Second, for an imposed pressure gradient, the flow velocity outside the tether decreases logarithmically as a function of the radial distance from the tether while the flow inside the tether is akin to Poiseuille flow. A tangential stress balance at the interface shows that for typical viscosities the flow is primarily outside the tether.  Third, by imposing normal stress balance at the oil-water interface, we obtain an analytical expression for the variation in the thread radius with distance along the thread and show that the thread radius decreases linearly with distance, with a slope that is proportional to the flow rate.  There is therefore an upper limit on the flow rate through a tether of a defined length. 

\textit{Tether formation:}
The surfactant monolayer at the oil-water interface leads to a non-zero spontaneous curvature $H_0$. The free energy cost for bending deformations of such a system is given by~\cite{p:Helfrich1973},
\begin{equation}
G = \int \left[ \sigma_{0} + 2 \kappa \left( H - H_{0} \right)^{2} + \bar{\kappa} K + \ldots \right] dA, \label{eq:Helfrich}
\end{equation}
where $\sigma_{0}$ is the free energy per unit area of the oil-water interface at its spontaneous curvature, $H = \left( c_{1} + c_{2} \right)/2$ and $K = \left( c_{1}c_{2} \right)$, the mean and Gaussian curvatures respectively, with $c_{1}$ and $c_{2}$ being the two principal curvatures of the surface. The elastic moduli corresponding to bending and saddle-splay deformations are given by $\kappa$ and $\bar{\kappa}$ respectively, and $dA$ denotes the surface area element. 

For a cylindrical tether of radius $r_{t}$ and length $\ell$, the local curvature is identical everywhere and the principal curvature along the long axis is zero, \textit{i.e.} (say) $c_{1}=0$ as the surface is flat. The principal curvature orthogonal to the long axis is the inverse of the tether radius, $c_{2} = 1/r_{t}$. Thus it follows that the Gaussian curvature, $K = c_{1} c_{2} = 0$, while the mean curvature $H = 1/\left( 2 r_{t} \right)$. Eq.~\ref{eq:Helfrich} can thus be written as 
\begin{equation}
G = 2 \pi \ell r_{t} \left[ \left(\sigma_{0} + 2 \kappa H^{2}_{0} \right) + \frac{\kappa}{2 r^{2}_{t}} - \frac{2 \kappa H_{0}}{r_{t}} \right]. \label{eq:Helfrich-expanded}
\end{equation}
Minimising the free energy functional w.r.t. $r_{t}$, $\frac{\partial G}{\partial r_{t}} = 0$ gives us the equilibrium tether radius~\cite{p:Prost2002}.
\begin{equation}
r_{t} = \sqrt{\frac{\kappa}{2 \tilde{\sigma}}},\label{eq:tether-radius}
\end{equation}
where $\tilde{\sigma} = \sigma_{0} + 2 \kappa H^{2}_{0}$ is the free energy per unit area of a planar oil-water interface. The tensile force on the tether is given by the minimisation of the free energy w.r.t. length $\ell$, \textit{i.e.} $\frac{\partial G}{\partial \ell}=0$,
\begin{equation}
f = 2 \pi \sqrt{2 \kappa \tilde{\sigma}} - 4 \pi \kappa H_{0}.~\label{eq:tether-force} 
\end{equation}

We contrast this with force required to pull a lipid tether in context of GUVs~\cite{p:Dommersness2005,p:Prost2002} where the tether length with a constant volume constraint is imposed. Since the thread is connected to two droplets, if the thread contracted at a rate such that flow to the reservoirs is insignificant, then a constant volume constraint is appropriate. However for the experimental situation outlined here the volume of the reservoirs is large compared to the tether dimension and in quasi-equilibrium conditions allows for free exchange of fluid and lipids resulting in the tether diameter remaining constant. Therefore a constant tether volume constraint, applicable for GUVs is not applicable for this situation. 

It is interesting to note that an unconstrained minimisation of the free energy Eq.~\ref{eq:Helfrich-expanded} w.r.t the length $\ell$ yields the same threshold force as the one in which the volume constraint has been enforced. Important differences however arise for the different experimental conditions discussed in the imposed boundary conditions to obtain the velocity profile of fluid transport through tethers.

The pressure difference across the lipid tether can be easily computed by taking the derivative of the free energy in Eq.~\ref{eq:Helfrich-expanded} with respect to the tether volume $V = \pi r^{2}_{t} \ell$ while keeping the tether length $\ell$ fixed. Thus the pressure difference is given by~\cite{p:Dommersness2005}: 
\begin{equation}
\left[ \frac{\partial G}{\partial V} \right]_{\ell} = \Delta p = \left[ \frac{\tilde{\sigma}}{r_{t}} - \frac{1}{2} \frac{\kappa}{r^{3}_{t}} \right],\label{eq:pressure-across-tether}
\end{equation}
where $\tilde{\sigma} = \sigma_{0} + 2 \kappa H^{2}_{0}$. The equilibrium tether radius can be obtained by noting that at equilibrium the pressure difference across the tether is zero, \textit{i.e.} $\Delta p = 0$, giving the same expression of the tether radius as in Eq.~\ref{eq:tether-radius}. 

Ultralow interfacial tensions (ULIFT) are typically observed near the phase transition from an oil-in-water microemulsion ($H_{0} > 0$) to a water-in-oil microemulsion ($H_{0} < 0$). Sometimes a bicontinuous ‘middle phase’ with  ($H_{0} = 0$) is also observed (so-called because it is lies between an excess oil phase and an excess aqueous phase). Microemulsion formulations can be tuned by variation of the temperature, salt concentration or co-surfactant concentration~\cite{p:Mead1986}. The experimental interfacial tension, which for most techniques involves a quasi-planar interface ($H \approx 0$) and can therefore be identified with $\tilde{\sigma}$, which shows a sharp minimum in the region where the middle phase exists.  For the system reported in \cite{p:Bain2006, p:Bain2011}, which comprises heptane, brine and the anionic surfactant AOT, the phase inverts from water-in-oil to oil-in-water with increasing temperature or decreasing salinity. In the immediate vicinity of the phase transition point, we can posit that the intrinsic interfacial tension, $\sigma_{0}$, is independent of temperature or salinity while the spontaneous curvature varies linearly. Figure \ref{fig:Infinite_Cylinder} shows the variation in the equilibrium tether radius $r_{t}$, renormalised interfacial tension $\tilde{\sigma}$, mean curvature $H$, and tension $f$ as a function of spontaneous curvature $H_{0}$ (scaled by the equilibrium tether radius of a bilayer $\sqrt{\kappa/2 \sigma_{0}}$, for which $H_{0}=0$). The interfacial tension $\tilde{\sigma}$, equilibrium tether radius $r_{t}$ and hence the mean curvature $H$, are all reflection symmetric under $H_{0} \rightarrow - H_{0}$. However as seen in Fig.~\ref{fig:Infinite_Cylinder} this is not true for the thread tension. As opposed to a lipid bilayer (having zero spontaneous curvature), an extra term appears in the expression for the thread tension, which tells us that the thread's free energy is lower under conditions favouring a microemulsion of the same sense. The experimentally measured thread tensions~\cite{p:Bain2011} decrease monotonically with increasing temperature ($H_{0}$ increasing from negative to positive values) in agreement with the predictions of our model.

\begin{figure}
\includegraphics[scale=0.375]{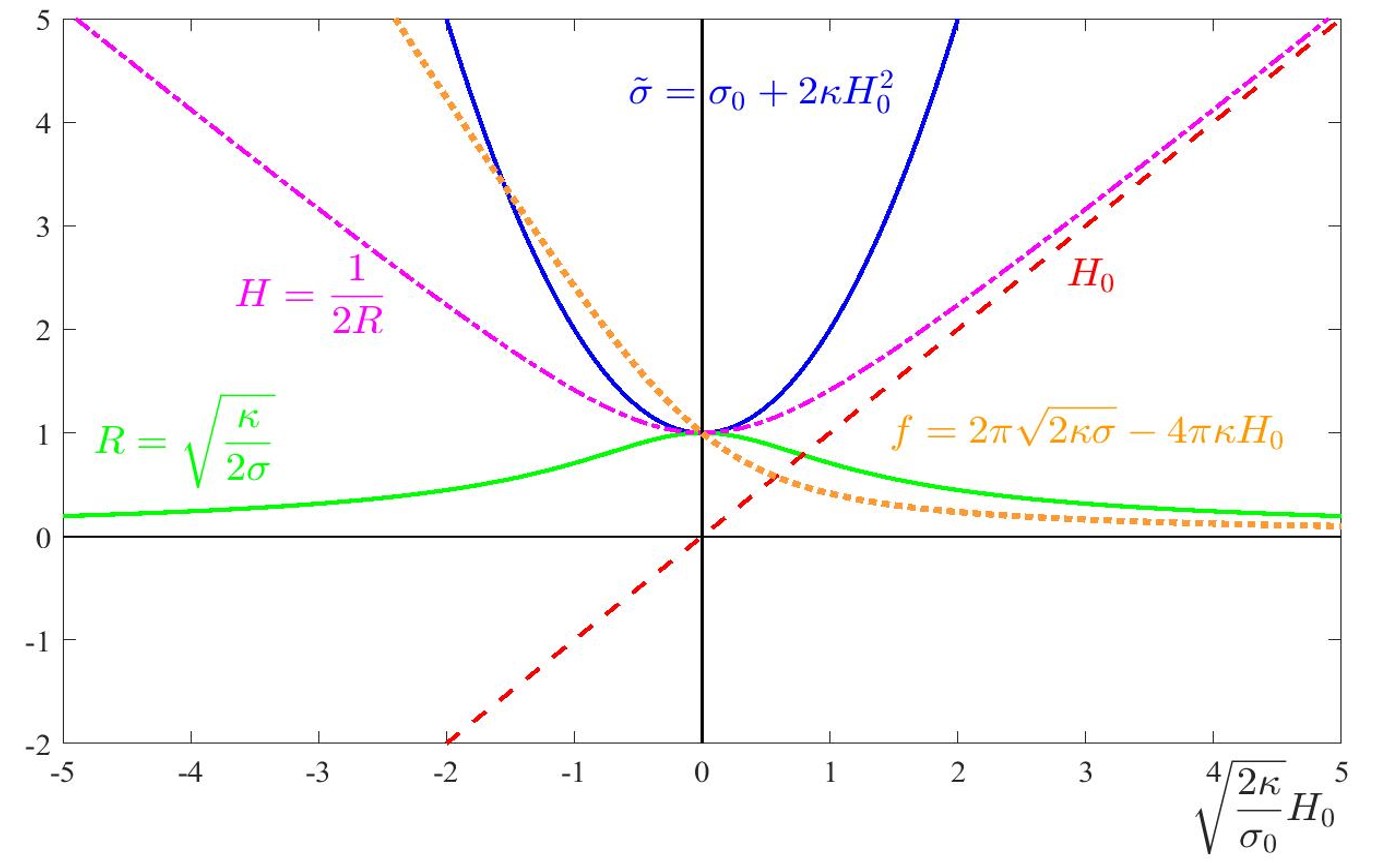}
\caption{Figure showing the variation of surface tension $\tilde{\sigma}$ (blue solid line), the radius of the tether $R$ (green solid line), the mean curvature of the tether $H$ (dash-dotted magenta line), and spontaneous curvature of the oil-water interface $H_{0}$ (red-dashed line) with its non-dimensionalised value. All functions are scaled by their value at the inversion point at which $H_{0} = 0$, and $\sigma = \sigma_{0}$.}\label{fig:monolayer-tether}
\end{figure}

\textit{Fluid flow inside and outside tethers} 
The flow through the flexible tether poses a challenging problem since the tether shape and fluid flow are intrinsically coupled. For the formulation used in this paper low Reynolds number hydrodynamics $Re << 1$ is assumed~\cite{b:Happel1983}. Further we assume that the tether radius $r_{t}$ is small compared to the radius of the terminal droplets. This is a valid approximation since we consider micron sized droplets while the tether radius is $\approx 50 \textrm{nm}$~\cite{p:Bain2011}.

We compute the steady-state flow profile by solving the Stoke's problem both inside and outside the tether and matching stresses at the boundaries. Consider the tether geometry shown in Fig.~\ref{fig:Infinite_Cylinder} with the $z$ component of the velocity varying along the radial direction $r$. The convention we follow is that the radiation pressure acting on the droplet is $-ve$, thus $p(z) < 0$. Consider in the schematic picture that the pressure on the left chamber of the system is zero, \textit{i.e} $p_{0}=0$. We choose our origin at the junction between the droplet and the tether. The equilibrium radius at this point $z=0$ having zero internal pressure, \textit{i.e.} $p(z=0)=0$, is given by $R_{0}= \sqrt{\frac{\kappa}{2 \tilde{\sigma}}}$. Since the droplet diameter is large we neglect the Laplace pressure difference between the droplet and the external fluid. Thus for Stokes flow outside the cylinder 
\begin{equation}
\eta_{o} \nabla^{2} v^{o}_{z}(r) = 0, \label{eq:Stokes-external} 
\end{equation}
where $\eta_{o}$ is the viscosity of the external fluid (water) and $v^{o}_{z}(r)$ the external fluid velocity at a point $r$. Assuming cylindrical symmetry, Eq.~\ref{eq:Stokes-external} admits a solution $v^{o}_{z}(r) = A \ln \ r + B$, with $r$ being the radial distance from the center of the tether and $A$ and $B$ are constants determined by the boundary conditions. The boundary conditions for the external flow are $v^{o}_{z}(r=r_{t}) = v_{s}$, where $v_{s}$ is the surface velocity at the tether wall and $v^{o}_{z}(r=L) \approx 0$. The second boundary condition is an approximation in order to bypass the Stokes problem~\cite{b:Landau1959}. For a tether within a closed cell, a recirculatory flow will be established on a length scale which will be the smaller of the distance to the confining walls or the tether length. Typical microfluidic cells have depths of $O(100 \mu m)$ and typical tether lengths are $O(10 \mu m)$; in either case the value of $L >> r_{t}$. Solving for the flow outside using the above boundary conditions lead to the standard result for flow around a moving cylinder
\begin{equation}
v^{o}_{z}(r) = \frac{v_{s}}{\ln \left[\frac{r_{t}}{L} \right]} \ln \left[\frac{r}{L} \right]. \label{eq:velocity-external-solution}
\end{equation}

Similarly the flow inside the tether has the Poiseuille form 
\begin{equation}
v^{i}_{z}(r) = v_{s} - \frac{1}{4 \eta_{i}} \left( \frac{dp}{dz} \right) \left( r^{2}_{t} - r^{2} \right),\label{eq:velocity-internal-solution}
\end{equation}
where $dp/dz$ is the pressure gradient acting along the axis of the tether connecting the two reservoirs, and $\eta_{i}$ is the viscosity of the inner fluid (oil). Tangential stress balance at the interface requires that
\begin{equation}
\eta_{i} \frac{\partial v^{i}_{z}(r)}{\partial r} = \eta_{o} \frac{\partial v^{o}_{z}(r)}{\partial r}. \label{eq:tangential-stress}
\end{equation}
Eq.~\ref{eq:tangential-stress} leads to a consistency condition of the surface velocity $v_{s}$:
\begin{equation}
v_{s} = \frac{1}{\eta_{o}} \frac{r^{2}_{t}}{2} \left( \frac{dp}{dz} \right) \ln \left[ \frac{r_{t}}{L} \right]. \label{eq:vs-solution}
\end{equation}
Note that $v_{s}$ is independent of $\eta_{i}$. 

\begin{figure}
\includegraphics[scale=0.16]{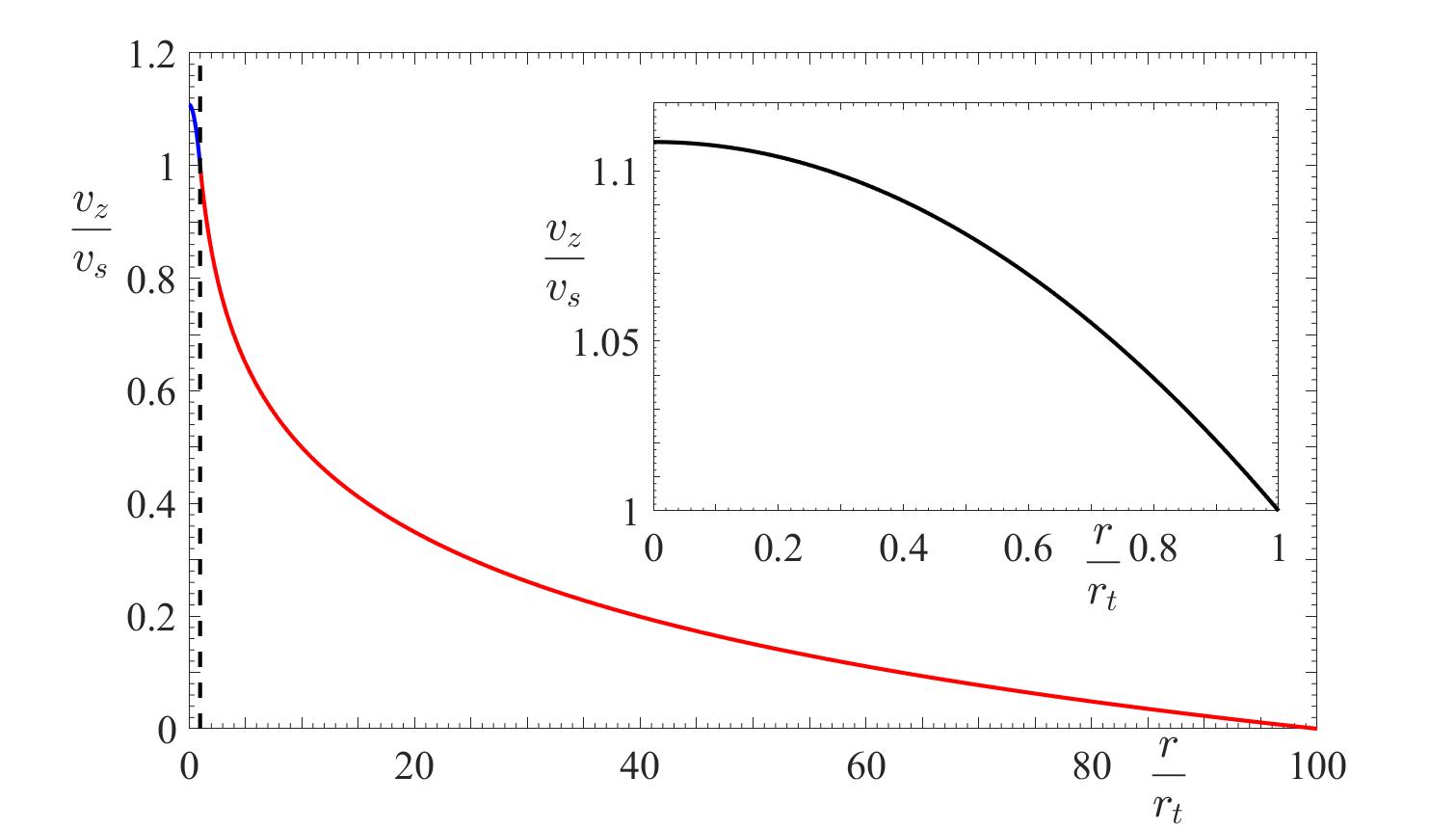}
\caption{Fluid velocity $v_{z}(r)$ as a function of the radial distance $r$ for $\frac{\eta_{i}}{\eta_{o}} = 1$. The velocity has been non-dimensionalised by scaling it with wall velocity $v_{s}$ ($v_s = v_{z}(r = r_{t})$) while the non-dimensionalised  radial distance is obtained by scaling $r$ by the tether radius $r_{t}$. The fluid velocity is zero at the walls $r = L$. In this figure $L = 100 r_{t}$. Inset shows an enlarged view of the non-dimensionalised fluid velocity profile inside the pipe.}
\label{fig:Velocity_profile}
\end{figure}
Fig.~\ref{fig:Velocity_profile} shows the velocity profile of the fluid flowing through the tether for $\frac{\eta_{i}}{\eta_{o}} = 1$, \textit{i.e.} given by Eq.~\ref{eq:velocity-external-solution} and ~\ref{eq:velocity-internal-solution}, outside and inside the tether with the velocity at the boundary being $v_{s}$. Fig.~\ref{fig:Velocity_profile} shows that the flow is mostly limited to the outside of the tether.

The full flow problem is complicated by the fact that $v_{s}$, $r_{t}$ and $dp/dz$ are all functions of $z$, for non-negligible flow rates $V_{f}$. We now determine the full steady-state velocity profile as a function of the flow rate.

The volumetric flow rate through the tether is given by,
\begin{equation}
V_{f} = \int^{r_{t}}_{0} 2 \pi r v^{i}_{z}(r) dr, \label{eq:Volumetric-flow}
\end{equation}
Utilising the velocity profile obtained in Eq.~\ref{eq:velocity-internal-solution} we have
\begin{equation}
V_{f} = \pi r^{2}_{t}(z) v_{s} - \frac{\pi}{8 \eta_{i}} \left(\frac{dp}{dz} \right) r^{4}_{t}(z), \label{eq:Volumetric-flow-exp}
\end{equation}

Substituting Eq.~\ref{eq:vs-solution} into Eq.~\ref{eq:Volumetric-flow-exp} we have 
\begin{equation}
V_{f} = \frac{\pi r^{4}_{t}(z)}{2 \eta_{o}} \left( \frac{dp}{dz} \right) \ln \left[\frac{r_{t}(z)}{L} \right] - \frac{\pi}{8 \eta_{i}} \left( \frac{dp}{dz} \right) r^{4}_{t}(z), \label{eq:Volumetric-flow-final}
\end{equation}
Inverting this relation we get an expression of the pressure gradient $\frac{dp}{dz}$ along the tether axis. 
\begin{equation}
\frac{dp}{dz} = - \frac{8 V_{f} \eta_{i}}{\pi r^{4}_{t}(z)} \left(1 - \frac{4 \eta_{i}}{\eta_{o}} \ln \left[\frac{r_{t}(z)}{L} \right] \right)^{-1}, \label{eq:pressure-gradient}
\end{equation}
The normal stress balance at the oil-water interface requires $dp/dz$ to be equal to the pressure gradient obtained by differentiating the pressure acting across the tether in Eq.~\ref{eq:pressure-across-tether} w.r.t $z$. 
\begin{equation}
\frac{dp}{dz} = - \left( \frac{\sigma}{r^{2}_{t}(z)} - \frac{3 \kappa}{2 r^{4}_{t}(z)} \right) \left( \frac{dr_{t}(z)}{dz} \right), \label{eq:pressure-gradient-mechanical}
\end{equation}
Eliminating the pressure gradient term between Eq.~\ref{eq:pressure-gradient} and Eq.~\ref{eq:pressure-gradient-mechanical} leads us to an equation for $r_{t}(z)$,
\begin{equation}
\frac{8 \eta_{i} V_{f}}{\pi} = \left[r^{2}_{t}(z) \tilde{\sigma} - \frac{3}{2} \kappa \right]  \left[ 1 - \frac{4 \eta_{i}}{\eta_{o}} \ln \left[ \frac{r_{t}(z)}{L} \right] \right] \left( \frac{d r_{t}(z)}{dz} \right). \label{eq:tether-radius-narrowing}
\end{equation}

Note that $4 \frac{\eta_{i}}{\eta_{o}} \ln[\frac{r_{t}(z)}{L}] >> 1$ implying that the dissipation in the outer fluid has a greater contribution than the inner fluid. Further note that $\sigma r^{2}_{t}(z) - \frac{3}{2} \kappa$ varies between $-\kappa$ and $-\frac{3}{2} \kappa$, corresponding to the equilibrium tether radius and zero tether radius as a function of length $\ell$. Thus Eq.~\ref{eq:tether-radius-narrowing} can be approximated as 
\begin{equation}
\ln \left[ \frac{r_{t}(z)}{L} \right] \frac{d r_{t}(z)/L}{dz} \simeq \frac{4 \eta_{o} V_{f}}{3 \pi \kappa L}. \label{eq:tether-thin-approx}
\end{equation}
Eq.~\ref{eq:tether-thin-approx} can be rewritten as
\begin{equation}
\frac{d}{dz} \left[ \frac{r_{t}(z)}{L} \ln \left(\frac{r_{t}(z)}{L} \right) - \frac{r_{t}(z)}{L} \right] = \frac{4 \eta_{o} V_{f}}{3 \pi \kappa L}. \label{eq:tether-thin-approx-preint}
\end{equation}
which can be integrated easily. Since $\lvert r_{t}(z) \ln \left[\frac{r_{t}(z)}{L} \right] \rvert >> r_{t}(z)$ the expression can be simplified to obtain 
\begin{equation}
r_{t}(z) \ln \left[ \frac{r_{t}(z)}{L} \right] \approxeq \frac{4 \eta_{o} V_{f}}{3 \pi \kappa} z + r_{t}(0) \ln \left[ \frac{r_{t}(0)}{L} \right]. \label{eq:tether-shape}
\end{equation}

Eq.~\ref{eq:tether-shape} gives an approximate solution for the shape of the tube. Note that $r_{t}(z) \rightarrow 0$ when the right hand side of Eq.~\ref{eq:tether-shape} is zero. This implies 
\begin{equation}
z_{\rm max} = \frac{3 \pi \kappa}{4 \eta_{o} V_{f}} r_{t}(0) \ln \left[ \frac{L}{r_{t}(0)} \right]. \label{eq:tether-length-lim}
\end{equation}
Plugging in values $V_{f} \approx 10^{-20} m^{3} s^{-1}$ \cite{p:Bain2011}, $\kappa \approx k_{B} T$~\cite{p:Muenier1994}, $\eta_{o} \approx 1 mPas$, $L/r_{t}(0) \approx 10^{3}$, and initial tether radius $R(0) = 50 \mathrm{nm}$ we have $z \approx 325 \mu m$. This sets a fundamental limit on the length of a nanotether in a network. Alternatively Eq.~\ref{eq:tether-length-lim} can be inverted to obtain a limit on the volumetric flow rate $V_{f}$ for a fixed tether length.

\begin{figure}
\includegraphics[scale=0.165]{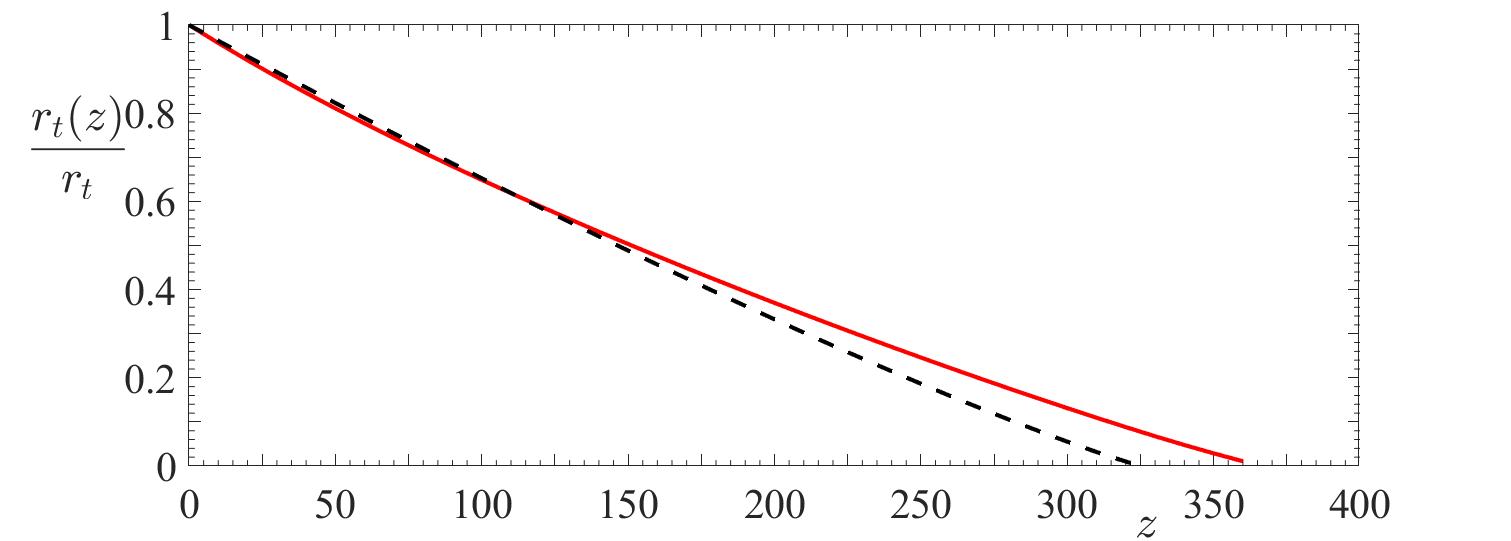}
\caption{The variation of the tether radius $r_{t}(z)$ scaled by the equilibrium tether radius $r_{t}$ as a function of the axial distance $z$ (measured in $\mu m$), obtained by numerically solving Eq.~\ref{eq:tether-radius-narrowing} (red solid line), and using the analytical approximation given by Eq.~\ref{eq:tether-shape} (black dashed line). For parameter values used to generate the plot see text.}
\label{fig:tether-narrowing}
\end{figure}

Figure \ref{fig:tether-narrowing} plots $r_{t}(z)$ as a function of $z$ for an initial tether radius of $50 nm$, derived from the full expression in Eq.~\ref{eq:tether-radius-narrowing}, with the approximation in Eq.~\ref{eq:tether-shape} shown as a dashed line.  We note that to a good approximation, $r_{t}$ decreases linearly with $z$.  The behaviour of $r_{t}$ as $r_{t} \rightarrow 0$ and the value of $z_{\rm max}$ in Eq.~\ref{eq:tether-length-lim} are only approximate since the assumptions leading to the derivation of Eq.~\ref{eq:tether-radius-narrowing} break down as the tether radius approaches zero.

\textit{Conclusions} 
Oil droplets in oil-in-water emulsions can be manipulated with focused laser beams (optical tweezers) when the interfacial tension is reduced by surfactants to sufficiently low values (comparable to the force constant of the optical traps). We have shown previously that when a single droplet is extended under pulling by two laser beams a configuration of two droplets connected by a single invisible nanothread of oil results. The radius of the tether, $r_{t}$, can be computed from the Helfrich Hamiltonian and is given by $r_{t} = \sqrt{\frac{\kappa}{2 \tilde{\sigma}}}$ where $\tilde{\kappa}$ is the bending modulus and the renormalized interfacial tension $\tilde{\sigma} = \sigma_{0} + 2 \kappa H^{2}_{0}$, $\sigma_{0}$ is the curvature-independent interfacial tension and $H_{0}$ is the spontaneous curvature of the interface.  We have considered how the mechanical properties of the thread vary in the vicinity of the phase inversion temperature (PIT) between a water-in-oil microemulsion and an oil-in-water microemulsion. Whereas $r_{t}$ and $\tilde{\sigma}$ are symmetric with respect to the PIT, the thread tension decreases monotonically with increasing spontaneous curvature, in agreement with experiment~\cite{p:Bain2011}, and contrary to a model that neglects changing spontaneous curvature \cite{p:Bain2011}.

Variation in the powers of the two laser traps holding the terminal droplets leads to a flow of oil through the connecting tether.   The flow profile is a combination of Poiseuille flow within the tether and ``flow around a rigid cylinder'' outside the tether. The velocity of the oil-water interface is found from tangential stress balance.  For typical ratios of viscosities of water and oil $\eta_{oil}/\eta_{water} > 1/3$, the flow is predominantly in the external fluid.  Thus transport of oil from one droplet to the other is mostly by motion of the whole oil thread, with interface being created at one droplet and destroyed at the other, rather than by flow through the thread itself.  The speed at which the oil-water interface moves is independent of the viscosity of the oil.  The normal stress balance on the o/w interface leads to the thread radius, $r_{t}$, decreasing monotonically with distance in the direction of flow.  An analytical expression is found $r_{t}(z)$ which is shown to be linear to a good approximation except near the point the point where $r_{t} \rightarrow 0$, where the approximations in the model break down. For typical values of interfacial tension, bending modulus and volumetric flow rate reported in the literature, the initial thread radius is $\sim 50 nm$ and would decrease to zero after $325 \mu m$.  There is therefore an upper limit on the volumetric flow rate that can be achieved by optical pumping through tethers of finite length. 

We hope that our work will inspire future experimental work to test the validity of this prediction.

\textit{Acknowledgements:} 
We thank EPSRC for funding support via grant $\mathrm{EP/I013377/1}$, Durham University for computational facilities, and Alex Lubansky and Jonny Taylor for helpful discussions.

\end{document}